\documentclass{webofc}
\usepackage[varg]{txfonts}   
\usepackage{dsfont}
\begin{document}
\title{A perturbative approach to the confinement-deconfinement\\ phase transition\footnote{Based on various works in collaboration with Julien Serreau, Matthieu Tissier and Nicol\'as Wschebor.}}
%
%

\author{\firstname{Urko} \lastname{Reinosa}\inst{1}\fnsep\thanks{\email{urko.reinosa@polytechnique.edu}}
}

\institute{CPHT, Ecole Polytechnique, CNRS, Universit\'e Paris-Saclay, F-91128 Palaiseau, France. 
          }

\abstract{%
We summarize recent progress in describing the confinement-deconfinement transition from a novel perturbative approach.}
\maketitle

The low energy regime of Quantum Chromodynamics is not accessible to standard perturbative methods  based on the Faddeev-Popov gauge fixing procedure, because the corresponding running coupling becomes large at these scales and even diverges at a finite scale known as $\Lambda_{\rm QCD}$. To remedy these shortcomings, various non-perturbative approaches have been invented, from lattice QCD to non-perturbative continuum methods, and much progress has been achieved over the years. Here, we would like to report on yet another possible route to the low energy regime of QCD. It is based on the speculation that perturbation theory might be applicable after all, provided one takes into account modifications of the Faddeev-Popov procedure. These modifications are needed because, in most practical cases, the gauge fixing is ambiguous in the infrared due to the presence of so-called Gribov copies. Various ways of dealing with the Gribov ambiguity in the Landau gauge have been proposed in the literature but we shall focus more specifically on the proposal of \cite{Serreau:2012cg}, based on the addition of a gluon mass term to the usual Landau gauge-fixed action.\footnote{We refer to \cite{Dudal08} and references therein for details on the so-called (refined) Gribov-Zwanziger approach.} This modified action remains renormalizable and, interestingly, some of the corresponding RG trajectories yield a running coupling that remains bounded and effectively small, even at low energies \cite{Tissier_10} which justifies the use of perturbation theory in this regime. In particular, at one-loop, the approach has been shown to reproduce lattice results for correlation functions to a good accuracy.

We shall further test the validity of this novel perturbative approach by studying the phase structure of QCD in the heavy-quark limit, for which the lattice has provided a wealth of data to compare with, including the case of a real chemical potential. In this regime, the relevant transition is the confinement-deconfinement transition, which we study by evaluating the Polyakov loop effective potential $V(\ell,\bar\ell)$. The appropriate extremization of the latter, see below, allows us to access the actual values of the Polyakov loops ($P$ denotes the path-oredering and $\bar P$ the anti-path ordering)
\begin{equation}\label{eq:PL}
\ell\equiv\frac 13{\rm tr}\,\left\langle P \exp\left(i g\int_0^\beta \!d\tau A_0^at^a\right)\right\rangle \quad {\rm and} \quad \bar \ell\equiv\frac 13{\rm tr}\,\left\langle \bar P \exp\left(-i g\int_0^\beta \!d\tau A_0^at^a\right) \right\rangle,
\end{equation}
as functions of the temperature $T=1/\beta$. These quantities are directly related to the free energies of a static quark and a static anti-quark respectively, $\ell=e^{-\beta F_q}$ and $\bar\ell=e^{-\beta F_{\bar q}}$, and are thus order parameters for the confinement-deconfinement transition. In fact, they are strict order parameters only in the limit of infinite quark masses, where they probe the breaking of the center symmetry of Yang-Mills theory at finite temperature, but they remain nevertheless good signatures of the transition for large but finite quark masses. 

\section{Polyakov loop effective potential at one-loop order}
\label{sec-1}
By definition, the Polyakov loop effective potential is a gauge-invariant function. However, computing it perturbatively requires one to fix the gauge. Here it is important to realize that not all gauges are equivalent within a given approximation method. In particular, if one aims at obtaining a clear signal of the transition, which is governed by center symmetry in the pure glue case, one should work in a gauge that does not break this symmetry explicitly. One possibility is to use the so-called Landau-DeWitt gauge, see for instance \cite{Braun:2007bx}, which leads to the following Faddeev-Popov contribution to the QCD action:
\begin{equation}\label{eq:FP}
    S_{\rm FP}=\int_x \Big\{(\bar D_\mu \bar c)^a(D_\mu c)^a+ih^a(\bar D_\mu a_\mu)^a\Big\}.
\end{equation}
Here $\smash{a_\mu\equiv A_\mu-\bar A_\mu}$ with $\bar A_\mu$ a given background field configuration and $\bar D_\mu^{ab}\equiv\delta^{ab}\partial_\mu+gf^{acb}\bar A^c_\mu$ is the corresponding covariant derivative. Since the Landau-DeWitt gauge suffers from the Gribov ambiguity, we modify the Faddeev-Popov action by adding the term
\begin{equation}\label{eq:Sm}
  S_m=\int_x \frac 12 m^2 a_\mu^a a_\mu^a\,,
\end{equation}
along the lines of \cite{Serreau:2012cg,Tissier_10,Reinosa:2014ooa}. In the SU(3) case, the value of the mass that allows to reproduce lattice data for the correlators at zero temperature is roughly $m\simeq 510\,{\rm MeV}$ and we shall consider this value in what follows.

A priori, the background $\bar A$ is arbitrary. However, within a given approximation scheme such as the perturbative scheme considered here, it might be convenient to choose a specific $\bar A$ in order to unveil certain properties of the system. In particular, at finite temperature $T$, it is natural to choose temporal and uniform background fields and, without loss of generality, one can take them along the Cartan subalgebra of $su(3)$: $g\bar A_\mu(x)=T\delta_{\mu0}(r_3\lambda_3+r_8\lambda_8)/2$. Moreover, in order to keep a good handle on center symmetry, it is convenient to choose so-called self-consistent backgrounds, such that $\bar A=\langle A\rangle_{\bar A}$ where $\smash{\langle...\rangle_{\bar A}}$ denotes the expectation value in the Landau-DeWitt gauge (including the additional mass term) with background $\bar A$. It can be shown that such backgrounds play the role of order parameters for center symmetry in the pure Yang-Mills case \cite{Reinosa:2015gxn} and that they are obtained by minimizing a center-symmetric potential (see below for the appropriate extremization procedure in the presence of quarks), the so-called background effective potential $V(r_3,r_8)$. Its one-loop expression (including quarks) in the modified gauge-fixing defined by Eqs.~(\ref{eq:FP}) and (\ref{eq:Sm}) can be found in \cite{Reinosa:2015oua}. Correspondingly, one also obtains the tree-level expressions for the Polyakov loops in terms of the background components:
\begin{equation} \label{eq:tree}
\ell=\frac{1}{3}\left[e^{-i\frac{r_8}{\sqrt{3}}}+2e^{i\frac{r_8}{2\sqrt{3}}}\cos(r_3/2)\right] \quad {\rm and} \quad \bar\ell=\frac{1}{3}\left[e^{i\frac{r_8}{\sqrt{3}}}+2e^{-i\frac{r_8}{2\sqrt{3}}}\cos(r_3/2)\right].
\end{equation}
By inverting this mapping and by plugging it back into the one-loop expression for $V(r_3,r_8)$, one arrives at a one-loop expression for the Polyakov loop effective potential. More precisely, one obtains $V(\ell,\bar\ell)=V_{\rm gauge}(\ell,\bar\ell)+\sum_f V_f(\ell,\bar\ell)$, with
\begin{equation}
V_{\rm f}(\ell,\bar\ell)=-\frac{T}{\pi^2}\int_0^\infty dq\,q^2\left\{\ln\Big[1+3\ell\,e^{-\beta(\varepsilon^f_q-\mu)}+3\bar\ell\,e^{-2\beta(\varepsilon^f_q-\mu)}+e^{-3\beta(\varepsilon^f_q-\mu)}\Big]+(\mu\to-\mu)\right\}
\end{equation}
and
\begin{eqnarray}
V_{\rm gauge}(\ell,\bar\ell) & \!\!\!\!\!\!\!=\!\!\!\!\!\!\! & \frac{T}{\pi^2}\int_0^\infty dq\,q^2\,\left\{\frac{3}{2}\ln\Big[1+e^{-8\beta\varepsilon_q}-(9\ell\bar\ell-1)(e^{-\beta\varepsilon_q}+e^{-7\beta\varepsilon_q})\right.\nonumber\\
& & \hspace{2.5cm}+\,(27\ell^3+27\bar\ell^3-27\ell\bar\ell+1)(e^{-2\beta\varepsilon_q}+e^{-6\beta\varepsilon_q})\nonumber\\
& & \hspace{2.5cm}-\,(81\ell^2\bar\ell^2-27\ell\bar\ell+2)(e^{-3\beta\varepsilon_q}+e^{-5\beta\varepsilon_q})\nonumber\\
& & \hspace{2.5cm}\left.+\,(162\ell^2\bar\ell^2-54\ell^3-54\bar\ell^3+18\ell\bar\ell-2)e^{-4\beta\varepsilon_q}\Big]-\frac{1}{2}(m\to 0)\right\},\nonumber\\
\end{eqnarray}
where $\smash{\varepsilon_q\equiv\sqrt{q^2+m^2}}$ and $\smash{\varepsilon_q^f\equiv\sqrt{q^2+M^2_f}}$, with $M_f$ the mass of the quark of flavour $f$. In what follows, we discuss the phase structure that emerges from these one-loop expressions, both at imaginary and at real chemical potential, and compare it with lattice predictions.

\section{Phase structure at imaginary chemical potential}
The case of an imaginary chemical potential has attracted a lot of attention in recent years for the fermion determinant that appears under the QCD functional integral is real and positive. This means that there is no sign problem and lattice simulations provide accurate results in this case. Although this situation does not correspond to physical QCD, one hopes at gaining some knowledge about the structural properties of the theory and even at extracting some information about the case of a real chemical potential, through analytic continuation.

\begin{figure}[h]
\centering
\sidecaption
\includegraphics[width=4cm,clip]{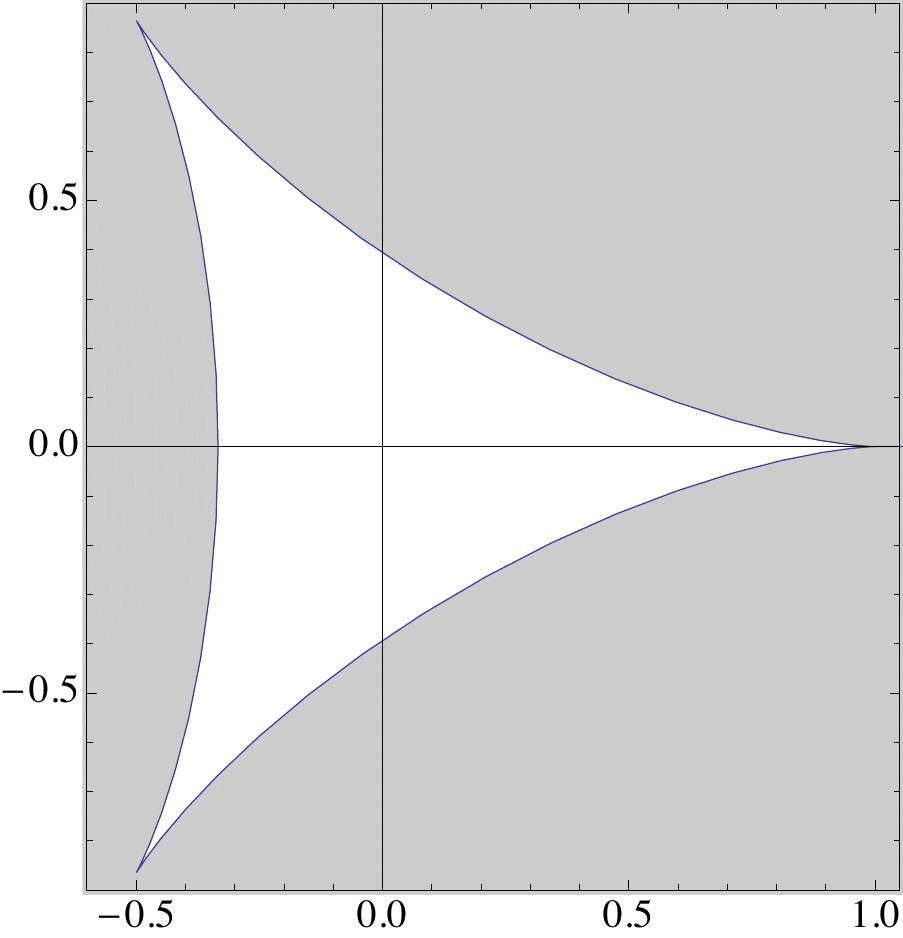}
\caption{Allowed values (white region) for the Polyakov loop in the complex plane for $\ell$, in the case of an imaginary chemical potential.}
\label{fig-region_im}       
\end{figure}

The fermion determinant being real, it is simple to argue that the Polyakov loops defined in Eq.~(\ref{eq:PL}) should be complex conjugate of each other, as expectation values of complex conjugate quantities in the presence of a real weight under the functional integral. This means that we can restrict the study of $V(\ell,\bar\ell)$ to pairs $(\ell,\bar\ell)$ that belong to $\Sigma\equiv\{(u,v)\in\mathds{C}^2\,|\,v=u^*\}$. In other words, this means that we can consider $V(\ell)\equiv V(\ell,\bar\ell=\ell^*)$. This fact, together with the tree-level expressions (\ref{eq:tree}), is compatible with our findings in \cite{Reinosa:2015oua}, where in the case of an imaginary chemical potential, we argued that the background components $r_3$ and $r_8$ could be taken real. We note however that, even though the pair $(r_3,r_8)$ can take any value in $\mathds{R}\times\mathds{R}$, $\ell$ and $\bar\ell$ are only allowed to take certain values in the complex plane, see figure \ref{fig-region_im}. We shall thus restrict the analysis of $V(\ell)$ to this region.

As already mentioned, the fermion determinant is not only real but positive. This in turn allows to argue that the physical value (\ref{eq:PL}) of the Polyakov loop is obtained at the absolute minimum of $V(\ell)$, more details will be given elsewhere. With this rule in mind, we can now discuss the phase structure that emerges from our one-loop potential.

\pagebreak

Let us first start with the infinite quark mass limit, that is the pure Yang-Mills case. We find a first order phase transition at a temperature $T_d\simeq 185\,{\rm MeV}$, see \cite{Reinosa:2014ooa}, the unique, $Z_3$-symmetric minimum of $V(\ell)$ at $\smash{\ell=0}$ for $T<T_d$, turning into a triplet of $Z_3$-breaking minima with $\smash{\ell\neq 0}$ for $T>T_d$. The order of the transition is in agreement with the lattice predictions, even though the value for $T_d$ is somewhat below the lattice value $T_d\simeq 270\,{\rm MeV}$, see for instance \cite{Lucini:2012gg}. However, one should keep in mind that the present calculation is just a one-loop result and one needs to assess how large the corrections are. We have computed the two-loop corrections to the background effective potential and we have again found a first order phase transition with $T_d\simeq 254\,{\rm Mev}$, see Ref.~\cite{Reinosa:2015gxn}, a value much closer to the lattice result or to results from functional methods \cite{Fister:2013bh}.

\begin{figure}[h]
\centering
\sidecaption
\includegraphics[width=5cm,clip]{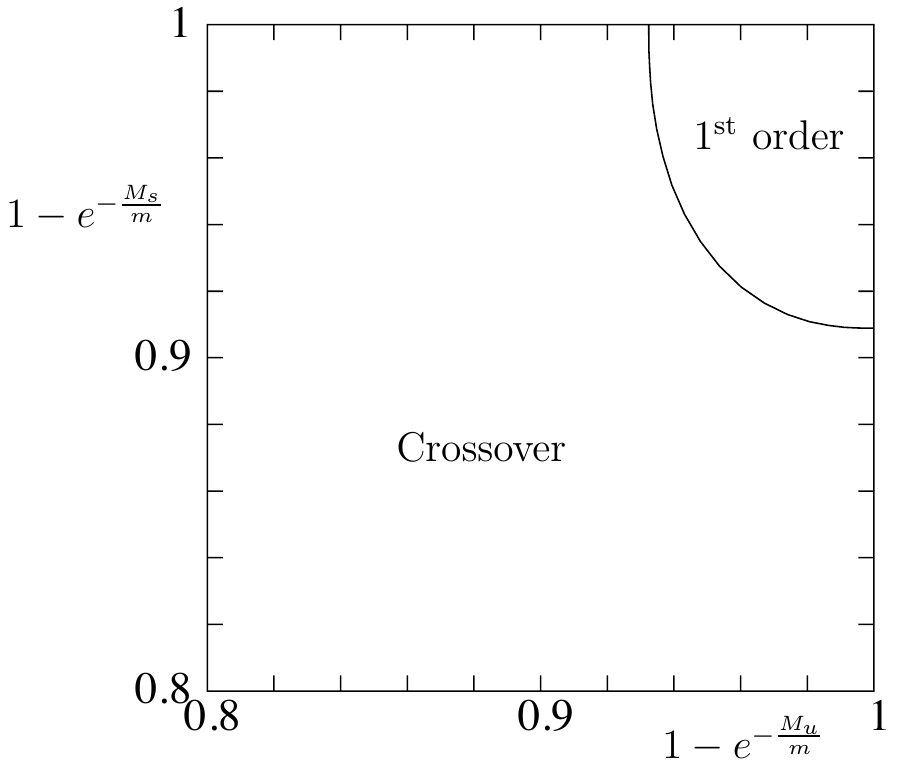}
\caption{Top right corner of the Columbia plot as obtained from our perturbative approach at one-loop order.}
\label{fig-Columbia}       
\end{figure}

Including the quarks, first at zero chemical potential, we find that the nature of the transition depends on the values of the quark masses. For large enough $u$, $d$ and $s$ quark masses, the transition remains of first order. But as the masses are lowered, one encounters a line of second order phase transitions in the $(\smash{M_u=M_d},M_s)$ plane (the so-called Columbia plot), see figure \ref{fig-Columbia}. For even smaller masses, the transition becomes a crossover.
The agreement with lattice results is not only qualitative but also quantitative as shown in table~\ref{tab-1} where we show the ratio of the critical mass to the critical temperature for $N_f=1,2,3$ degenerate flavours. It is also worth noting that, at this order of approximation, this ratio does not dependent on the specific value of the gluon mass parameter $m$.
\begin{table}[h]
\centering
\caption{Ratio of the critical quark mass to the critical temperature for $N_f=1,2,3$ degenerate flavours, as obtained in the present approach (second column) and compared to results from the lattice \cite{Fromm:2011qi} (third column) and a matrix model \cite{Kashiwa:2012wa} (fourth column).}
\label{tab-1}       
  \begin{tabular}{|c|c|c|c|}
\hline
$\,N_f\,$&$\,M_c/T_c\,$&$\left(M_c/T_c\right)^{\rm latt.}$&$\left(M_c/T_c\right)^{\rm matr.}$\\
\hline
1&6.74&7.22(5)&8.04\\
\hline
2&7.59&7.91(5)&8.85\\
\hline
3&8.07&8.32(5)&9.33\\
\hline
  \end{tabular}
\end{table}
We mention finally that, as long as $\smash{\mu=0}$, $\ell$ remains real (and positive), in agreement with its interpretation in terms of the quark free energy. Since $\smash{\bar\ell=\ell^*}$, this implies also that $\smash{\ell=\bar\ell}$, in line with the fact that the quark and anti-quark free energies need to be equal since charge conjugation is manifest at $\mu=0$. 

\begin{figure}[h]
\centering
\sidecaption
\includegraphics[width=6cm,clip]{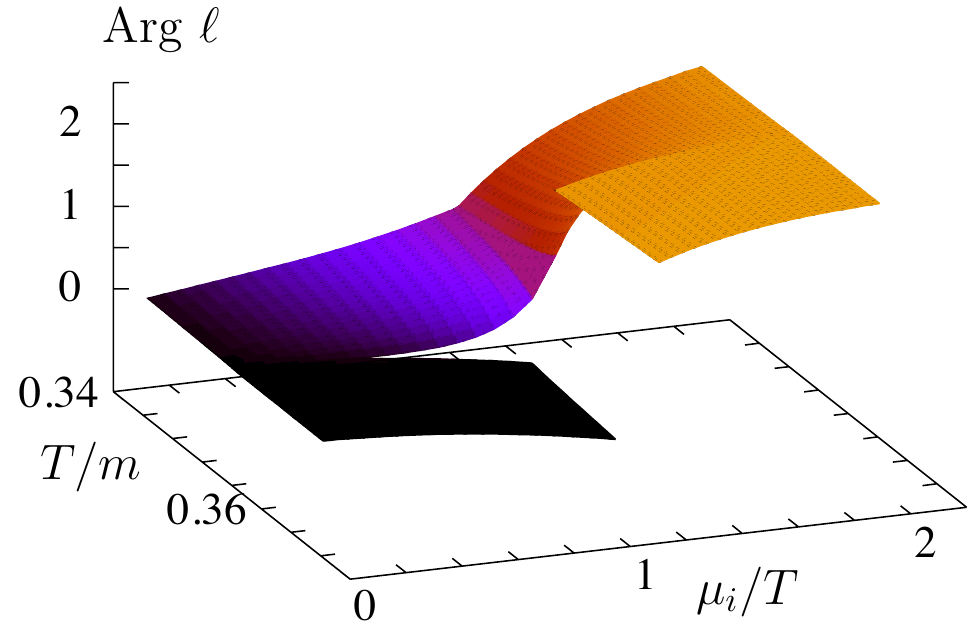}
\caption{Argument of the Polyakov loop as a function of the temperature and imaginary chemical potential. For large enough temperatures, the phase undergoes a first order phase transition at $\mu_i/T=\pi/3$.}
\label{fig-RW}       
\end{figure}

As we turn on the chemical potential along the imaginary axis $i\mathds{R}$, we observe that the Polyakov loop $\ell$ acquires a phase which rotates as $\mu_i\equiv{\rm Im}\,\mu$ is increased. If the temperature is large enough, this occurs with a first order jump at $\mu_i/T=\pi/3$. We have thus recovered the well known Roberge-Weiss transition, see figure \ref{fig-RW}.
 The dependence of the phase structure on the quark masses is again in qualitative agreement with that observed on the lattice. In particular, we observe that the critical points of the Columbia plot at $\mu=0$ enter the imaginary chemical potential phase diagram as the mass is decreased below the critical mass. By decreasing the mass even further, two of such critical points end up meeting at a tricritical point whose scaling is essentially governed by the corresponding mean-field exponent:
\begin{equation}
\label{eq:tricsca}
 \frac{M_c(\mu)}{T_c(\mu)}= \frac{M_{\rm tric.}}{T_{\rm tric.}}+K\left[\left(\frac{\pi}{3}\right)^2+\left(\frac{\mu}{T}\right)^2\right]^{2/5},
\end{equation}
as observed also on the lattice \cite{deForcrand:2010he}. The agreement with the lattice also extends to non-universal quantities such as the ratio $M_{\rm tric.}/T_{\rm tric.}$ or the prefactor $K$. We find $M_{\rm tric.}/T_{\rm tric.}=6.15$ and $K=1.85$, to be compared with the lattice result of Ref.~\cite{Fromm:2011qi}, $(M_{\rm tric.}/T_{\rm tric.})^{\rm latt.}=6.66$ and $K^{\rm latt.}=1.55$ for $3$ degenerate quark flavours.

\section{Phase structure at real chemical potential}
In the case of a real chemical potential, the fermion determinant becomes complex. This is at the root of the sign problem in lattice QCD and, as we discuss below, leads to a (milder) sign problem in continuum approaches.

\begin{figure}[h]
\centering
\sidecaption
\includegraphics[width=5cm,clip]{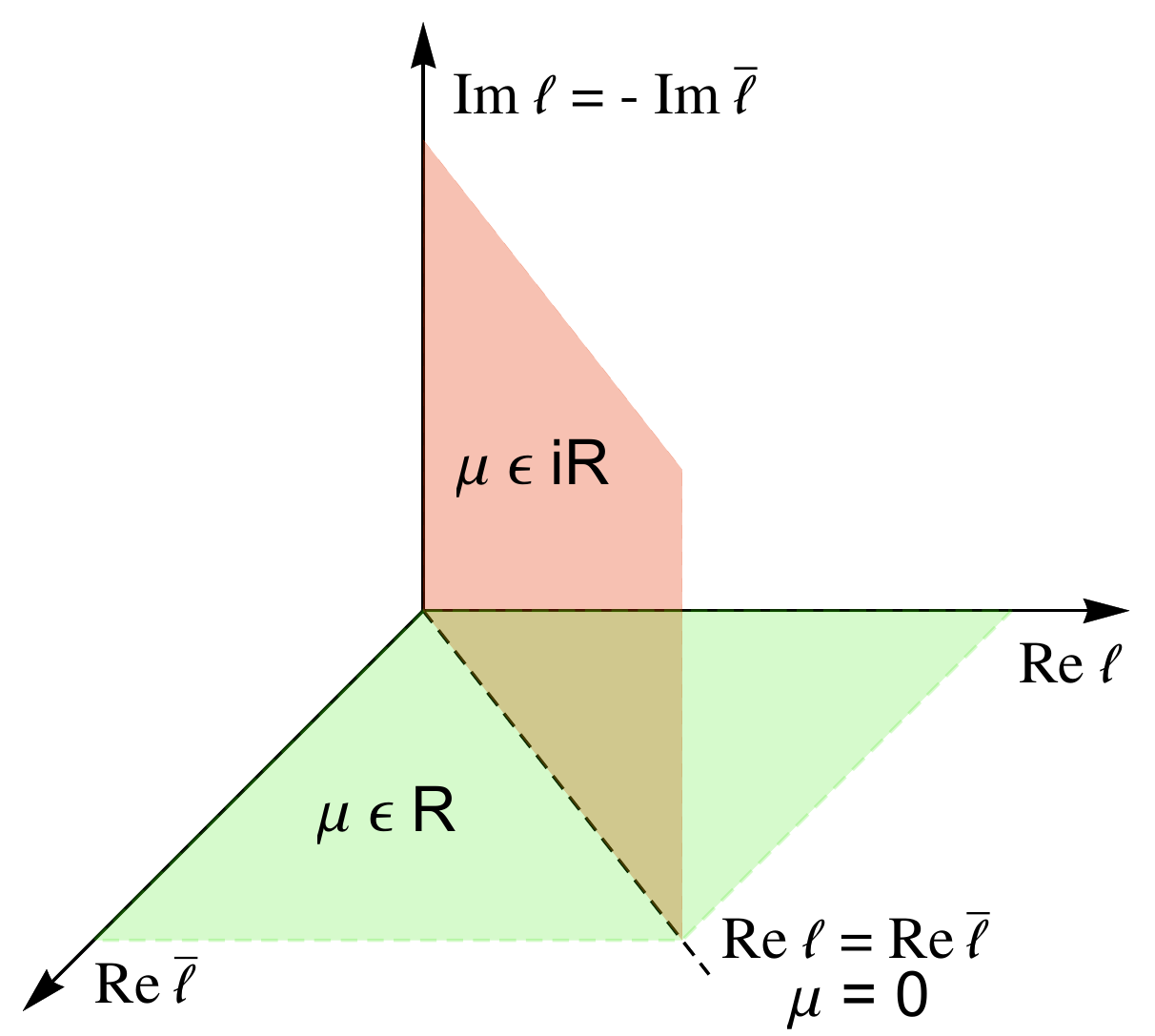}
\caption{The two different planes where the Polyakov loop effective potential should be considered depending on whether $\mu\in i\mathds{R}$ or $\mu\in\mathds{R}$. These two planes are submanifolds of $\mathds{C}\times\mathds{C}$, but for the sake of representing them in a three dimensional figure, we considered the space $\{(\ell,\bar\ell)\in\mathds{C}^2\,|\,{\rm Im}\,\ell=-{\rm Im}\,\bar \ell\}$.}
\label{fig-im-re}       
\end{figure}

First of all, it is not true anymore that $\ell$ and $\bar\ell$ are complex conjugate of each other and it is not consistent to restrict the analysis of the Polyakov potential to $V(\ell,\bar\ell=\ell^*)$. In fact, using some symmetries under the QCD functional integral, one shows that $\ell$ and $\bar\ell$ are both real and independent a priori. This means that the analysis of $V(\ell,\bar\ell)$ can be restricted to pairs $(\ell,\bar\ell)\in\mathds{R}\times\mathds{R}$. This change of paradigm from the case of imaginary chemical potential is represented in figure~\ref{fig-im-re}. Once again, it is consistent with our findings in the case of the background field effective potential $V(r_3,r_8)$ for which we found that, in the case of a real chemical potential, $r_3$ needed to be taken real, while $r_8$ needed to be taken imaginary. We mention once more that, if $r_3$ and $r_8$ are free to take any value in $\mathds{R}$ and $i\mathds{R}$ respectively, this is not the case for the pair $(\ell,\bar\ell)$ which explores only a subregion of $\mathds{R}\times\mathds{R}$ to which one should restrict the analysis of $V(\ell,\bar\ell)$.

The previous discussion does not really qualify as a sign problem. In fact, it does not originate in the fermion determinant not being positive but only in the fermion determinant not being real. Moreover, it is clear what needs to be done, that is taking $\ell$ and $\bar\ell$ real (and independent a priori) as follows from some of the symmetries of the QCD action. We mention also that the same symmetries allow us to argue that $V(\ell,\bar\ell)$ is real if $\ell$ and $\bar\ell$ are taken real, a welcomed property if we expect to extract any meaningful thermodynamical information from the potential. 

The true sign problem reemerges instead in that it is now not completely clear how to extract the physics from the potential $V(\ell,\bar\ell)$. Even though the physical point of the system is certainly an extremum of $V(\ell,\bar\ell)$ for it corresponds to the limit of zero sources, it is not clear which extremum one should choose in the case where various extrema exist. In fact the characterization of the zero source limit as the absolute minimum of $V(\ell,\ell^*)$ in the case of an imaginary chemical potential does not apply to the case of a real chemical potential. This is rooted in the non-positivity of the fermion determinant and has thus the same origin than the lattice sign problem.

\begin{figure}[h]
\centering
\includegraphics[width=5cm,clip]{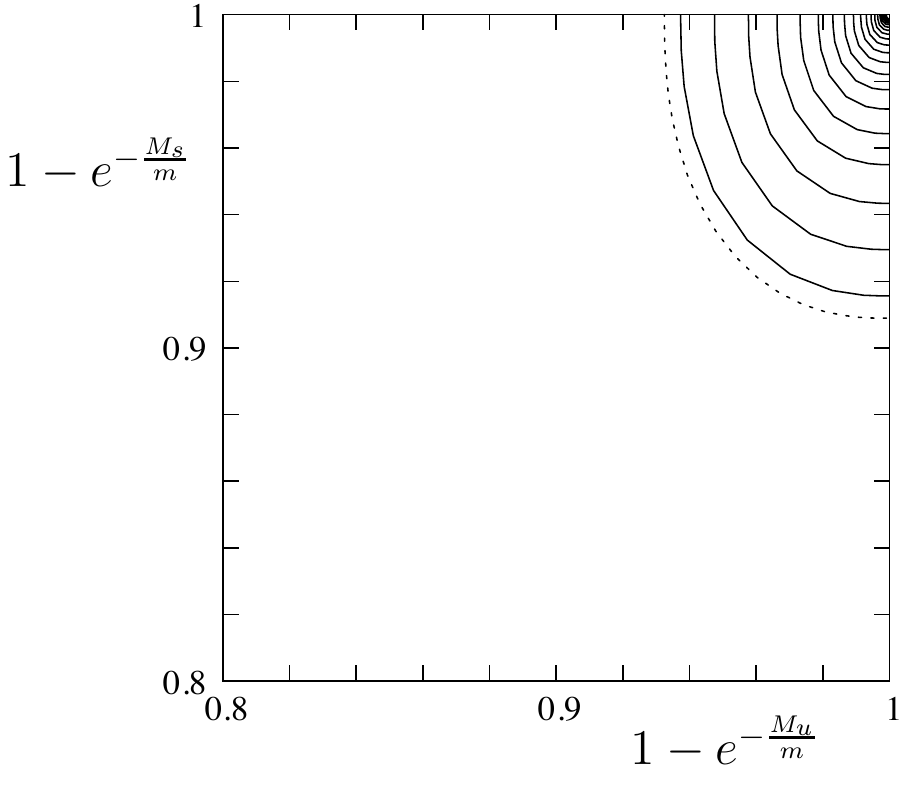}\qquad\qquad\includegraphics[width=6cm,clip]{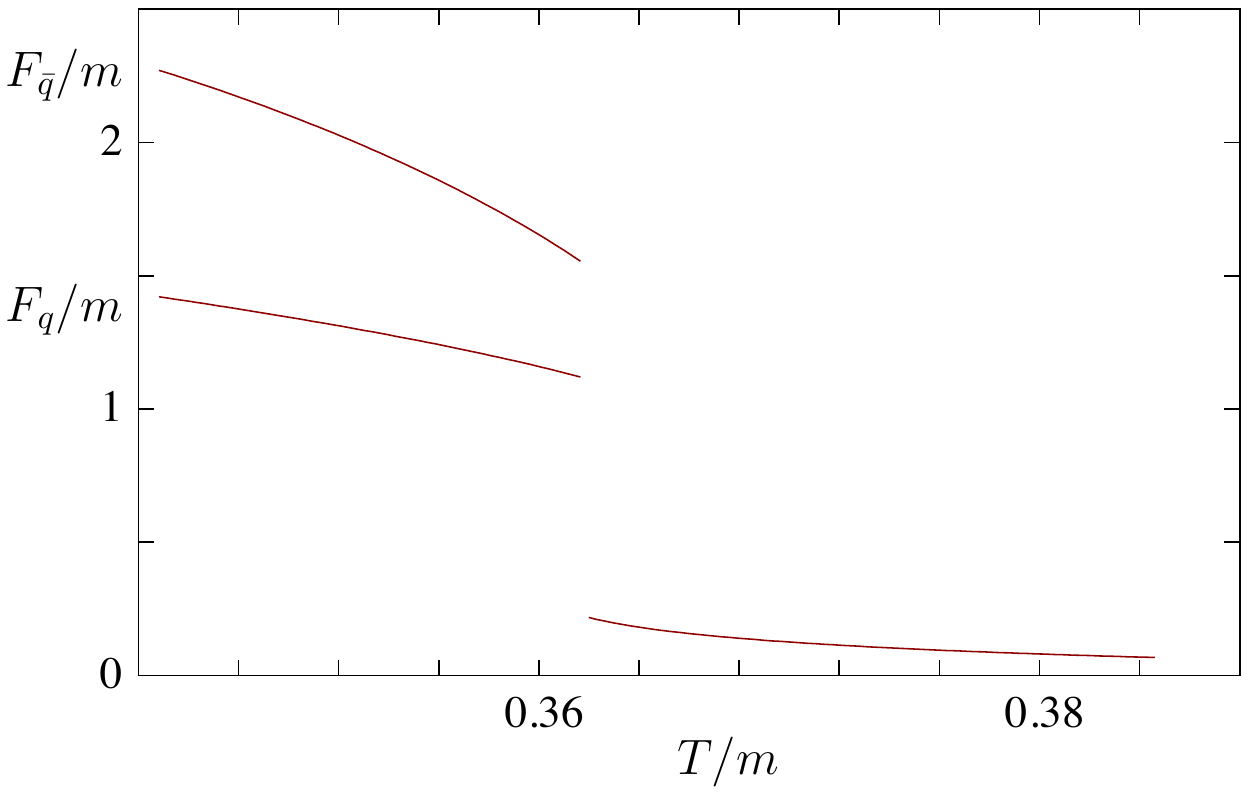}
\caption{Left: Columbia plot at non-zero real $\mu$. Projected critical surface on the $\mu=0$ plane. Right: Quark and anti-quark free-energies as functions of the temperature and across the (first order) transition.}
\label{fig-Cmu}       
\end{figure}

To proceed, we thus need to choose a recipe for selecting the physical extremum. To gain some insight, we can go back to the $\mu=0$ case for which the physical extremum is characterized as a minimum in the vertical plane of figure \ref{fig-im-re} and lies at the intersection with the horizontal plane. Viewed from the perspective of this latter plane, it appears as a saddle point, actually the deepest saddle point. For any other $\mu\in\mathds{R}$, we then choose to identify the physical extremum with the deepest saddle point, but we emphasize that we do not have a rigorous mathematical justification of this choice, even though it sounds plausible in terms of the minimisation of the Gibbs energy among all possible extrema. Given this recipe, our result for the Columbia plot for increasing $\mu$ is depicted in the left plot of figure \ref{fig-Cmu}. We find that the critical surface is pushed towards larger masses, as is also observed on the lattice.

We mention finally that our choice of an imaginary component $r_8$ and thus of real and independent Polyakov loops $\ell$ and $\bar\ell$ is crucial if one wants to obtain different quark and anti-quark free-energies, see the right plot of figure \ref{fig-Cmu}, in line with the explicit breaking of charge conjugation symmetry that a non-zero $\mu$ implies. In other approches, $r_8$ is taken equal to $0$ without much justification, which leads to $\ell=\bar\ell$ and thus to $F_q=F_{\bar q}$.

%
%
%

\end{document}